\documentclass[twocolumn,preprintnumbers,pre,floatfix,superscriptaddress,amsmath,amssymb]{revtex4}
\usepackage{epsfig}
\usepackage{subfigure}
\usepackage{amsmath}
\usepackage{amssymb}
\usepackage{setspace}
\usepackage{graphicx}
\usepackage{dcolumn}
\usepackage{bm}
\input epsf

\begin{document}

\author{Juan G. Restrepo}
\email{juanga@math.umd.edu} \affiliation{ Institute for Research in
Electronics and Applied Physics, University of Maryland, College
Park, Maryland 20742, USA } \affiliation{ Department of Mathematics,
University of Maryland, College Park, Maryland 20742, USA }

\author{Edward Ott}
\affiliation{ Institute for Research in Electronics and Applied
Physics, University of Maryland, College Park, Maryland 20742, USA }
\affiliation{ Department of Physics and Department of Electrical and
Computer Engineering, University of Maryland, College Park, Maryland
20742, USA }

\author{Brian R. Hunt}
\affiliation{ Department of Mathematics, University of Maryland,
College Park, Maryland 20742, USA } \affiliation{ Institute for
Physical Science and Technology, University of Maryland, College
Park, Maryland 20742, USA}

\date{\today}

\begin{abstract}
The largest eigenvalue of the adjacency matrix of the networks is a
key quantity determining several important dynamical processes on
complex networks. Based on this fact, we present a quantitative,
objective characterization of the dynamical importance of network
nodes and links in terms of their effect on the largest eigenvalue.
We show how our characterization of the dynamical importance of
nodes can be affected by degree-degree correlations and network
community structure. We discuss how our characterization can be used
to optimize techniques for controlling certain network dynamical
processes and apply our results to real networks.
\end{abstract}

\title{Characterizing the dynamical importance of network nodes and links}


\maketitle

In recent years, there has been much interest in the study of the
structure of networks arising from real world systems, of dynamical
processes taking place on networks, and of how network structure
impacts such dynamics \cite{newman1}. Remarkably, the largest
eigenvalue of the network adjacency matrix (which we denote
$\lambda$) has recently emerged as the key quantity determining many
important properties for the study of a variety of different
dynamical network processes. Some examples are the following: (i)
for a heterogeneous collection of chaotic and/or periodic dynamical
systems coupled by a network of connections, the critical coupling
strength \cite{onset} for the emergence of coherence is proportional
to $1/\lambda$ (ii) the critical disease contagion probability for
the onset of an epidemic \cite{wang} scales as $1/\lambda$; (iii) in
percolation on a network, the condition for the emergence of a giant
component also involves $\lambda$ \cite{footnote}. In addition to
these, there are other notable examples where $\lambda$ plays a
similar role \cite{may,siamlambda,survey}.

In many situations it might be desirable to control dynamical
processes that take place on networks. For example, in epidemic
spreading, one would like to increase the threshold for epidemic
transmission. In percolation, one might like to identify the key
nodes holding the network together and protect them (e.g., in the
transportation network or the internet) or disrupt them (e.g., in
the case of a terrorist network or pathogen protein network). Such
strategies would greatly benefit from a quantitative
characterization of the effect of the removal of the different nodes
or edges in the network. We will define the {\it dynamical
importance} of nodes and edges as the relative change in the largest
eigenvalue of the network adjacency matrix upon their removal. This
provides an objective quantification of the relative importance of
the different elements of the network that could potentially be used
to formulate control strategies for those network processes that are
governed by the largest eigenvalue of the network adjacency matrix.
We also will describe an efficient way to approximate the dynamical
importance.

We consider a network as a directed graph with $N$ nodes, and we
associate to it a $N\times N$ adjacency matrix whose elements
$A_{ij}$ are positive if there is a link going from node $i$ to node
$j$ with $i\neq j$ and zero otherwise ($A_{ii} \equiv 0$). We denote
the largest eigenvalue of $A$ by $\lambda$, where $Au =\lambda u$
and $v^T A = \lambda v^T$ with $u$ and $v$ denoting the right and
left eigenvectors of $A$. According to Perron's theorem
\cite{siamlambda}, of all the eigenvalues of $A$, the one with
largest magnitude is real and positive and the components of the
eigenvectors $u$ and $v$  all have the same sign (which we choose to
be positive). It is often the case that $\lambda$ is well separated
from the second largest eigenvalue.  We define the {\it dynamical
importance} of edge $i\to j$, $I_{ij}$, as the amount $-\Delta
\lambda_{ij}$ by which $\lambda$ decreases upon removal of the edge,
normalized by $\lambda$:
\begin{equation}
\addtolength{\belowdisplayskip}{-0.2cm}\addtolength{\abovedisplayskip}{-0.2cm}
 I_{ij} \equiv -\frac{\Delta\lambda_{ij}}{\lambda}.
\end{equation}
Similarly, the dynamical importance of node $k$ is defined in terms
of the amount $-\Delta \lambda_{k}$ by which $\lambda$ decreases
upon removal of the node (or equivalently removal of all edges into
and out of node $k$):
\begin{equation}\label{impor}
\addtolength{\belowdisplayskip}{-0.2cm}\addtolength{\abovedisplayskip}{-0.2cm}
I_{k} \equiv -\frac{\Delta\lambda_{k}}{\lambda}.
\end{equation}
\hspace{3mm}We will now use a perturbative analysis in order to
provide approximations $\hat I$ to the dynamical importance $I$ in
terms of $u$ and $v$. We first consider the importance of an edge
$I_{ij}$. Let us denote the matrix before the removal of the edge by
$A$ and after the removal by $ A + \Delta A$, the largest eigenvalue
of $A + \Delta A$ by $\lambda + \Delta\lambda$ and its corresponding
right eigenvector by $u + \Delta u$. For large matrices, it is
reasonable to assume that the removal of a link or node has a small
effect on the spectral properties of the network, so that $\Delta u$
and $\Delta \lambda$ are small. Left multiplying
\begin{equation}\label{leftmul}
\addtolength{\belowdisplayskip}{-0.2cm}\addtolength{\abovedisplayskip}{-0.2cm}
(A + \Delta A) (u + \Delta u) = (\lambda + \Delta \lambda)(u +\Delta
u)
\end{equation}
by $v^T$ and neglecting second order terms $v^T\Delta A \Delta u$
and $\Delta \lambda v^T\Delta u$, we obtain $ \Delta \lambda = v^T
\Delta A u/v^T u$. Upon removal of edge $i\to j$, the perturbation
matrix is $(\Delta A)_{lm} = -A_{ij}\delta_{il}\delta_{jm},$ and
therefore
\begin{equation}\label{Iij}
\addtolength{\belowdisplayskip}{-0.2cm}\addtolength{\abovedisplayskip}{-0.2cm}
\hat I_{ij} = \frac{A_{ij}v_i u_j} {\lambda v^T u}.
\end{equation}
We now examine the effect of removing node $k$. Upon its removal,
the perturbation matrix is given by $ (\Delta A)_{lm} =
-A_{lm}(\delta_{lk} + \delta_{mk}).$ However, in this case we cannot
assume $\Delta u$ is small as we did before, since $\Delta u_k =
-u_k$ (the left and right eigenvectors have zero $k^{th}$ entry
after the removal of node $k$). Therefore, we set $\Delta u = \delta
u - u_k \hat e_k$, where $\hat e_k$ is the unit vector for the $k$
component, and we assume $\delta u$ is small. Left multiplying
Eq.~(\ref{leftmul}) by $v^T$ and neglecting second order terms
$v^T\Delta A \delta u$ and $\Delta \lambda v^T\delta u$, we obtain $
\Delta \lambda = (v^T\Delta A u -u_k v^T \Delta A \hat e_k)/(v^T u
-v_k u_k)$. Using the expression for $\Delta A$, we get $v^T\Delta A
u = -2\lambda u_k v_k$ and $u_k v^T \Delta A \hat e_k = \lambda u_k
v_k$. Considering the network to be large ($N\gg 1$), we assume $u_k
v_k \ll v^T u$ and obtain
\begin{equation}\label{I}
\addtolength{\belowdisplayskip}{-0.2cm}\addtolength{\abovedisplayskip}{-0.2cm}
\hat I_{k} = \frac{v_k u_k} {v^T u}.
\end{equation}

The commonly used {\it eigenvector centrality} \cite{bonacich} of
node $k$ is defined as the eigenvector component $u_k$. Although
closely related to it, $I_k$ and $I_{ij}$ take into account the
possible asymmetry of $A$ and are defined in such a way that they
quantify the relative change in $\lambda$ upon removal of the node
or link. For a review of other measures of node importance, see
\cite{newman1} and references therein.

We will now present examples of the dynamical importance of nodes in
simulated and real networks. We consider unweighted networks (i.e.,
the nonzero elements $A_{ij}$ are identically one), but we emphasize
that our considerations also apply to weighted networks. In
considering the simulated examples, we will try to mimic the often
complex structure of real world networks. This complexity is
evidenced by the fact that the degree distribution in a large number
of examples has been found to be highly heterogeneous (often
following a power law in the so-called scale-free networks
\cite{barabasi2}), where the {\it out-degree} and {\it in-degree}
are defined by $d_i^{out}=\sum_{j=1}^N A_{ij}$ and
$d_i^{in}=\sum_{j=1}^N A_{ji}$. In the case of an undirected network
$A = A^T$ and $d_i^{in} = d_i^{out}\equiv d_i$. The `degree
distribution' $P(d^{in},d^{out})$ is defined as the probability that
a randomly chosen node has degrees $d^{in}$ and $d^{out}$ [in the
undirected case we write $P(d)$ to denote the corresponding degree
distribution]. Furthermore, it has been observed that the degrees at
the ends of a randomly chosen edge often have positive or negative
correlations (referred to as {\it assortative} or {\it
disassortative} mixing by degree \cite{mixing}, respectively). For
example, the ratio
\begin{equation}
\addtolength{\belowdisplayskip}{-0.2cm}\addtolength{\abovedisplayskip}{-0.2cm}
\rho = \langle d_i^{in}d_j^{out}\rangle_e/\langle d_i\rangle_e^2,
\end{equation}
where $\langle\dots\rangle_e$ denotes an average over edges,
$\langle Q_{ij}\rangle_e \equiv
\sum_{i,j}A_{ij}Q_{ij}/\sum_{i,j}A_{ij}$, is larger (smaller) than
$1$ in assortative (disassortative) networks.

A mean field approximation, $A_{ij} \propto d_i^{out}d^{in}_j$,
yields $\rho = 1$, $u_i = d_i^{out}$, and $v_i = d_i^{in}$, and thus
$\hat I_k = d_k^{out}d_k^{in}/(\sum_{k=1}^N d_k^{out}d_k^{in})$. We
will denote this reference importance by $\hat I^0_k$. (For an
undirected network this is equivalent to ranking nodes by their
degree \cite{immunization}.)
\begin{figure}[h]
\centering \epsfig{file = 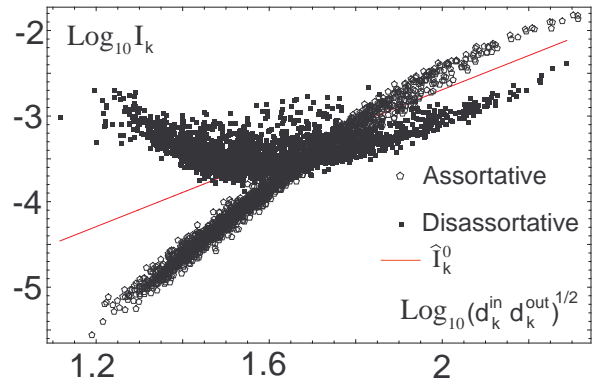, clip =  ,width=0.95\linewidth
}
\addtolength{\belowcaptionskip}{-0.3cm}\addtolength{\abovecaptionskip}{-0.5cm}
\caption{Node dynamical importance $I_k$ and $\hat I^0_k$ (solid
line) as a function of $\log_{10}(d_k^{in} d_k^{out})$ for (a) an
assortative network (open circles), and (b) a disassortative network
(boxes). } \label{cuafi}
\end{figure}

Our first examples consist of networks with positive and negative
degree-degree correlations. We choose (somewhat arbitrarily) to
examine networks in which the in- and out- degrees at each node are
uncorrelated, $P(d^{in},d^{out}) = P_{in}(d^{in})P_{out}(d^{out})$,
and have the same distribution, $P_{in}(d)=P_{out}(d)=\hat P(d)$.
The networks are generated by first prescribing a target degree
sequence $(d_i^{in},d_i^{out})$. In order to generate networks with
a power law degree distribution, $\hat P(d)\propto d^{-\gamma}$, we
will use, following \cite{chung}, the sequence of expected degrees
$c(i+i_0-1)^{-1/(\gamma-1)}$ for the in-degrees, and a random
permutation of this sequence for the out-degrees, where
$i=1,\dots,N$, and $c$ and $i_0$ are chosen to obtain a desired
maximum and average degree. From these sequences, the adjacency
matrix is constructed by setting $A_{ij} = 1$ for $i\neq j$ with
probability proportional to $d_i^{out}d_j^{in}$ and zero otherwise
($A_{ii} = 0$) (this is a generalization of the model in Chung {\it
et al.} \cite{chung}). Finally, the following (based on a
simplification of the method in Ref.~\cite{mixing}) is repeated
until the network has the desired amount of degree-degree
correlations as evidenced in the value of $\rho$: Two edges are
chosen at random, say connecting node $i$ to node $j$ and node $n$
to node $m$, and are replaced with two edges connecting node $i$ to
node $m$ and node $n$ to node $j$ if $s(d_n^{in}d^{out}_m +
d_i^{in}d^{out}_j-d_n^{in}d^{out}_j-d_i^{in}d^{out}_m) < 0$, and are
left alone otherwise.  Setting $s = +1$ or $-1$ we produce
assortative or disassortative networks, respectively.

We construct two different asymmetric networks of size $N = 2000$
following the algorithm above with $\gamma = 2.5$, and $c$, $i_0$
chosen such that $\langle d \rangle = 50$ and $d_{max} = 350$. For
networks (i) and (ii) we used $s=+1$ and $s=-1$, respectively, until
a desired value of $\rho$ was reached. This resulted in values of
$\rho$ of $1.6$ and $0.69$ for networks (i) and (ii), respectively.

\begin{figure}[h]
\centering
\epsfig{file = 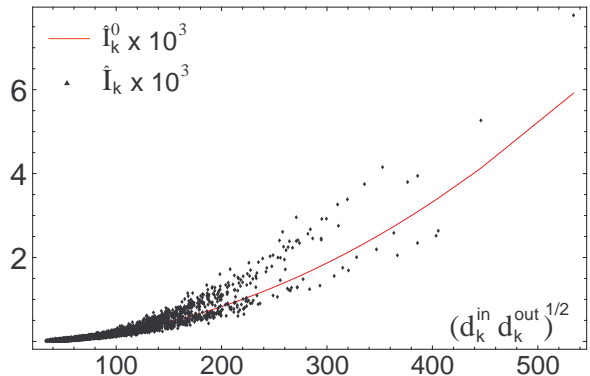, clip =  ,width=0.95\linewidth }                  
\addtolength{\belowcaptionskip}{-0.3cm}\addtolength{\abovecaptionskip}{-0.6cm}
\caption{Node dynamical importance $I_k$ for the two-community
network described in the text (stars) and the uncorrelated reference
$I^0_k$ (solid line).} \label{cuafiab}
\end{figure}

In Fig.~\ref{cuafi} we show on a logarithmic scale (base $10$) our
approximation to the node dynamical importance $\hat I_k$ versus
$\sqrt{d_k^{in}d_k^{out}}$ for networks (i) (open circles) and (ii)
(boxes), and the reference importance $\hat I^0_k$ (solid line). We
see that for the assortative network there is a rough monotonic
relation between $\hat I_k$ and $\sqrt{d_k^{in}d_k^{out}}$, while
for the disassortative case a functional relationship even less
clear and the nodes with low value of $\sqrt{d_k^{in}d_k^{out}}$
have importance comparable to the highly connected nodes (since, due
to the disassortativity, they act as bridges connecting the hubs).
In both cases $I^k$ and its approximation by $\hat I^k$ [Eq.
(\ref{I})] are essentially the same to within the size of the
plotted points in the figure (this will also apply to
Fig.~\ref{cuafiab}).

Our next example is motivated by the fact that it is sometimes
observed that real networks can be subdivided into more or less well
defined communities which have different statistics, and thus
potentially different dynamical importance. As a simple model of
such situation, we specify a division of the nodes in the network
into two groups of the same size, $A$ and $B$, ($A\bigcup B =
\{1,2,\dots,N\}$, $A\bigcap B = \emptyset$), and then we construct a
network following the steps above with $s=+1$, but only rewire the
edges if all the nodes in consideration belong all to group $A$
($i$, $j$, $n$, $m \in A$). The effect of this division is to create
a subnetwork (group $A$) with a correlation that is larger than that
for the whole network.

In Fig.~\ref{cuafiab} we show the node dynamical importance $I_k$
for this network (stars) and the uncorrelated reference $I^0_k$
(solid line). We see that the dynamical importance captures the
subdivision existing in the network. Nodes in the assortative region
$A$ are more dynamically important than nodes  with the same
connectivity $\sqrt{d_k^{in}d_k^{out}}$ outside of this region. This
shows that the node dynamical importance can depend on the
subdivision of the network into communities, and suggests that in
networks with strong community subdivision the node dynamical
importance could be weakly correlated with the degree.

\begin{figure}[h]
\centering
\epsfig{file = 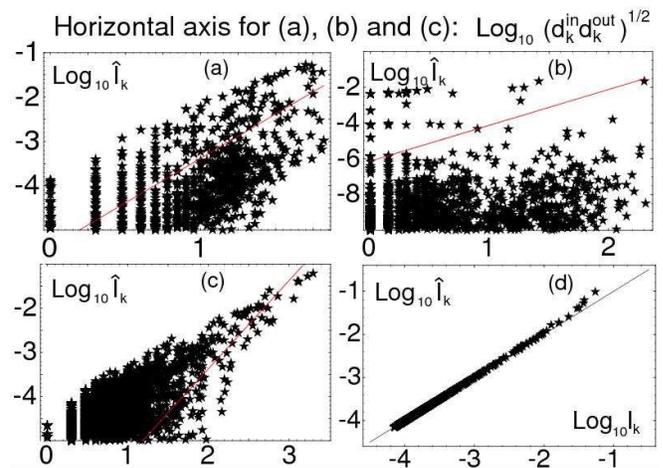, clip =  ,width=1.0\linewidth }    
\addtolength{\belowcaptionskip}{-0.7cm}\addtolength{\abovecaptionskip}{-0.6cm}
\caption{Logarithm of the dynamical importance $\hat I_k$ as a
function of the logarithm of $\sqrt{d_k^{in}d_k^{out}}$ for (a) the
yeast protein interaction network \cite{yeast1,yeast2}, (b) the Kiel
University email network \cite{email}, and (c) the internet (AS)
network \cite{caida}. Figure (d) shows $\hat I_k$ versus $I_k$ for
the AS network. The solid line is the identity.} \label{usair97}
\end{figure}

We will now consider the dynamical importance of the nodes in the
undirected yeast protein interaction network \cite{yeast1,yeast2}
($N = 2361$), the directed Kiel University email network
\cite{email} ($N = 64385$), and the undirected internet autonomous
systems (AS) network \cite{caida} ($N = 21885$). (Each one of these
networks is an incomplete sample of a larger network. For the
purpose of illustrating our ideas, we study the dynamical importance
of the reported nodes.) The dynamical importance of the nodes in
these three networks is shown as a function of $\sqrt{d^{in}
d^{out}}$ in a double logarithmic scale (base $10$) in Figs.~1
(a,b,c). The points were calculated from Eq.~(\ref{I}), except the
rightmost point in Fig. 1 (b), for which our assumption $v_ku_k \ll
v^Tu$ was not satisfied and for which we calculated $I_k$ directly
from the definition. Otherwise, the approximation $\hat I_k$ yielded
good results, except for a relatively small bias towards larger
values for $v_k u_k/v^T u \sim 0.1$. This is illustrated in Fig. 1
(d), which shows that, in spite of the deviation of $\hat I_k$ from
$I_k$, the relationship is still monotonic, and hence does not
affect the relative ranking of nodes.

A striking characteristic of the three plots is that, although there
is a correlation between dynamical importance and the connectivity
measured by $d^{in}d^{out}$, there are huge variations of importance
among nodes of comparable connectivity. In particular, for the
directed email network [Fig.~\ref{usair97} (b)], many poorly
connected nodes ($d^{in} d^{out} \sim 1-5$) have a dynamical
importance vastly larger than some of the most connected nodes
($d^{in} d^{out} \sim 10^4$). This suggests that, when enough
information about the network is available, the dynamical importance
of nodes should be a key element in the formulation of optimum
immunization strategies.
\begin{figure}[t]
\centering
\epsfig{file = 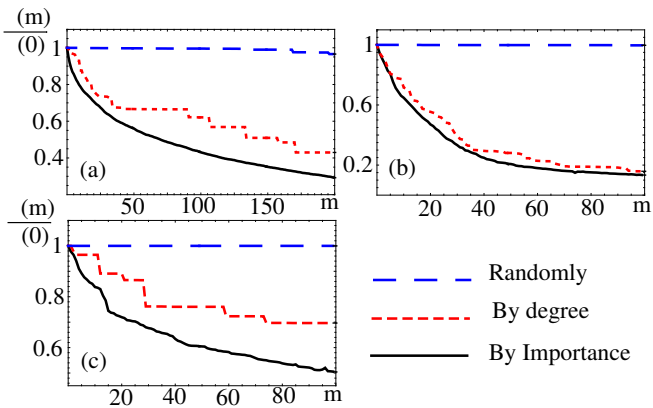, clip =  ,width=1.0\linewidth }              
\addtolength{\belowcaptionskip}{-0.6cm}
\addtolength{\abovecaptionskip}{-0.6cm} \caption{Largest eigenvalue
$\lambda(m)$ of the network resulting from removing $m$ nodes in
order of decreasing dynamical importance (solid lines), degree
(short dashed lines), and randomly (long dashed lines), for (a) the
yeast protein interaction network \cite{yeast1,yeast2}, (b) the
internet AS network \cite{caida}, and (c) the Kiel University email
network \cite{email}.} \label{fig4}
\end{figure}

We will now show how knowledge of the dynamical importance of nodes
can be used to optimally reduce $\lambda$ in order to control
various dynamical processes as discussed above. For the three
networks presented above, we successively remove either (i) the most
dynamically important nodes [as determined by Eq.~(\ref{I})], (ii)
the nodes with the highest value of $d^{in}d^{out}$, or (iii) random
nodes. (After removing a node, we recalculate the importance and the
degrees.) In Fig.~\ref{fig4} we show $\lambda(m)/\lambda(0)$ as a
function of $m$, where $m$ is the number of removed nodes and
$\lambda(m)$ is the largest eigenvalue of the resulting network. We
see that using the dynamical importance (solid lines) greatly
improves the results over using the degree (short dashed lines).
These two methods are, of course, much more efficient than removing
nodes randomly (long dashed lines).

We now offer some additional general comments. (i) The perturbation
technique used to obtain Eq.~(\ref{I}) also can be used to estimate
the increase of $\lambda$ upon addition of a new node and its links
or the addition or change in the weight of one link. If we are
allowed to add a new node with a prescribed number of in- and
out-links or to increase the weight of a link by a specified amount,
this approach can determine how to proceed so as to maximally
increase $\lambda$ (as in a case in which one would like to promote
synchronization \cite{onset}). (ii) The perturbation analysis can
also be applied to weighted networks that have some negative weight
links ($A_{ij} < 0$ for some $i,j$). In particular, if the number of
negative weight links is not too great, $\lambda$ is still real and
positive and the components of $u$ and $v$ are still positive. In
that case, Eq.~(\ref{Iij}) shows that removing such a link increases
$\lambda$ and, furthermore, one can use it to determine the best
negative weight link to remove if one wishes to most increase
$\lambda$. (iii) Another use of Eq.~(\ref{I}) might be to determine
from a given network a reduced simpler network with fewer nodes, but
almost the same dynamics (in the sense of having almost the same
$\lambda$). This could be done by successive removal of the nodes of
lowest $I_k$ (as in the idea of the `k-core' \cite{kcore}), and
offers a potential tool for facilitating the understanding of the
dynamics on a complex network. (iv) As compared to degree-based node
ranking, the approximation $\hat I_k$ of Eq.~(\ref{I}) requires
computation of the eigenvectors $u$ and $v$ and hence more complete
network information.

In conclusion, we have presented an objective, quantitative measure
of the dynamical importance of edges and nodes in a network. The
dynamical importance of a node or edge measures how the largest
eigenvalue, which controls various important dynamical processes in
networks, changes upon removal of the given node or edge. We have
shown how knowledge of the dynamical importance of nodes can be used
to optimize strategies to control dynamical processes dependent on
the largest eigenvalue of the adjacency matrix of the network.

E.O. acknowledges a conversation with Marc-Thorsten Hutt in which he
pointed out the possible relevance of components of eigenvectors
corresponding to the maximum eigenvalue of network matrices. This
work was supported by ONR (Physics), by the NSF (PHY 0456240 and DMS
0104-087), and by AFOSR.

\begin{spacing}{1.0}
\bibliographystyle{plain}

\end{spacing}

\end{document}